\documentclass[seceq]{ptptex}

\def\lsim{\mathrel{\rlap{\lower3pt\hbox{\hskip1pt$\sim$}}
     \raise1pt\hbox{$<$}}} 
\def\gsim{\mathrel{\rlap{\lower3pt\hbox{\hskip1pt$\sim$}}
     \raise1pt\hbox{$>$}}} 
\def\la{\langle}
\def\ra{\rangle}
\def\be{\begin{eqnarray}}
\def\ee{\end{eqnarray}}
\def\bi{\bibitem}\def\Tr{{\rm Tr}}
\def\B{\cal B}

\title{
Baryons and Vector Dominance in Holographic Dual QCD%
}

\author{
Mannque \textsc{Rho}%
}

\inst{
Institut de Physique Th\'eorique, Direction des Sciences de la
Mati\`ere, CEA-Saclay, F-91191 Gif-sur-Yvette, France}

\abst{
The infinite tower of vector mesons encoded in holographic dual
QCD bring drastic changes to the soliton structure of the nucleon.
The nucleon is given by a point-like instanton in 5D surrounded by a
vector meson cloud with the vector dominance restored by the
infinite tower. I discuss the possible relevance of this structure
in hot and dense hadronic matter.}

\begin{document}

\maketitle

\section{The Objective}
I would like to discuss in this talk a particular aspect of baryon
structure that arises from holographic dual QCD (hQCD for short)
that has an important ramification on some current issues of
nuclear/hadronic physics. My talk will be largely motivated by the
series of work~\cite{HRYY,HRYY-VD} I have done recently with my
string theory/particle physics colleagues in Korea. This work addressed
broadly two different issues. As nicely exposed in the recent talks
by my colleagues\cite{Yi,DKH}, to string theorists, it is already quite surprising that the notion of gravity/gauge duality can access
certain properties of the baryons at such an accuracy, say, $\sim
10\%$ level. Here string theory purports to first ascertain how well
it {\em postdicts} the baryon properties well described by QCD {\em
proper} or rather by its effective field theories and then to
address problems that go beyond the standard model, e.g., baryon
decay~\cite{baryondecay}, just to cite one.

What I am interested in here is quite different in nature. I would
like to see in what way string theory may provide us with something
that cannot, at present, be accessed  by QCD proper.

I would like to first describe what it is that we would like to
understand, describe what hQCD can actually do and then state what
needs to be done to enable hQCD to answer the question posed.
\section{The Origin of Hadron Mass}
One of the currently active researches in strong interaction physics
is to unravel how the ground-state hadrons, $\rho$, $\omega$, $p$
and $n$, that figure importantly in nuclear physics get most of
their masses, given that the masses of the basic constituents,
quarks, are tiny on the strong interaction scale. Numerous
experiments have been done at various laboratories in the world and
will continue in the upcoming facilities at CERN, GSI etc. in search
of evidence for the assumed mechanism of the mass generation. At the
intuitively simplest level, one may attribute the mass mostly to the
spontaneous breaking of chiral symmetry as prescribed by QCD. Taking
the $\rho$ (and $\omega$) as the prime example\footnote{The nucleon
mass is somewhat subtler, requiring more details, so I will not go
into it here. I advertise the monograph \cite{MR-CND2}for
discussions on this matter.}, this would suggest that the vector-meson mass $m_v$
should ``run" along the quark condensate $\la\bar{q}q\ra$ as the
latter slides due to a vacuum change. If correct, this
picture~\cite{MR-CND2} would predict that when nuclear matter is
heated as in heavy-ion collisions or compressed as in compact stars,
the mass would ``shed", and would go to zero in the chiral limit, as
the critical temperature or density is reached from below since QCD
dictates that the order parameter of chiral symmetry
$\la\bar{q}q\ra\rightarrow 0$. This feature is concisely captured in
what is called ``BR scaling."

The signal for BR scaling has been experimentally searched for with
dileptons as a snap-shot of the vector meson that propagates in
hot/dense medium created in heavy-ion collisions. The idea is to map
out the spectral functions of the vector meson as a function of
invariant mass and look for the ``dropping mass" effect below the
free-space peak. The results so far obtained are under debate with a
clear understanding still largely missing due to a plethora of
background processes that are unrelated to the chiral vacuum
properties being searched, but the consensus seems that the effect of
BR scaling is {\em not} visible in the putative spectral functions
so far extracted.

One hastily drawn conclusion was that ``BR scaling is ruled out by
Nature." This then brings one back to square zero: Where does $m_v$
come from? While this possibility cannot yet be excluded at present,
I will take the contrary view recently put forward~\cite{BHHRS} that
the dileptons measured in heavy-ion collisions, as they stand, carry
no direct information on chiral symmetry. I will then suggest how
hQCD with its infinite tower of vector mesons could confirm or
refute whether the notion of BR scaling is valid.
\section{Vector Dominance and Hidden Local Symmetry}
\subsection{Vector manifestation}
Recent developments indicate that at low energies below the chiral
scale, $\Lambda_\chi\sim 4\pi f_\pi$, strong interactions are
governed by hidden local symmetry theory denoted HLS$_\infty$
involving an infinite tower of flavor vector fields $v^{(k)}$, $k=1,
\cdots, \infty$ and that the dynamics of hadrons in medium with the
possibility of vector meson mass dropping to near zero requires that
local gauge symmetry be present. In this section, I will focus on
the lowest members $v=\rho$ (and $\omega$) in the presence of the
(pseudo-)Goldstone pion fields. This can be done by formally
integrating out all vector excitations of the tower except for the
lowest. Let me call the resulting theory HLS$_1$. This (nonabelian)
flavor gauge theory consisting of the Goldston pion field $\pi$ and
the gauge field $\rho$ is considered to be an effective field theory
of QCD valid at a scale much less than the chiral scale
$\lambda_\chi$ encoding the correct chiral symmetry pattern
$SU(N_f)_L\times SU(N_f)_R\rightarrow SU(N_f)_{L+R}$ and lends
itself to a systematic chiral perturbation theory including the
gauge degrees of freedom\cite{HY:PR}. The local gauge invariance in
this theory
allows one to easily do quantum loop calculations even when $m_v$
becomes comparable to that of the pion, which is zero in the chiral
limit, a situation characterizing our approach which is difficult,
if not impossible, to access in the unitary gauge used by other
workers, e.g., massive Yang-Mills or tensor formalism.

To the leading order in the derivative expansion and neglecting the
quark masses, HLS$_1$ has only three parameters, the gauge coupling
$g$, the pion decay constant $F_\pi$ and the decay constant of the
longitudinal component of the $\rho$ meson $F_\sigma$ or, more conveniently,  the ratio
$a=(F_\sigma/F_\pi)^2$. Matching the vector and axial-vector current
correlators of HLS$_1$ to the corresponding QCD correlators at a
matching scale $\Lambda_M \lsim \Lambda_\chi$ determines $g$,
$F_\pi$ and $a$ at $\Lambda_M$ in terms of the QCD variables
$\alpha_c$, $\la\bar{q}q\ra$, $\la G_{\mu\nu}^2\ra$ etc. at
$\Lambda_M$. Quantum one-loop calculations via renormalization group
equations (RGE) reveal that the theory possesses a variety of fixed
points. However, imposing that the vector and axial vector
correlators be equal when chiral symmetry is restored with
$\la\bar{q}q\ra=0$ (in the chiral limit), picks one particular fixed
point called ``vector manifestation (VM)"\cite{HY:PR}:
 \be
g^*=0, \ a^*=1.\label{fp}
 \ee
In this theory, there is nothing special about the behavior of the
parameter $F_\pi$ under RG flow other than that the physical
(on-shell) pion decay constant $f_\pi=F_\pi +\Delta$ (with $\Delta$
pion-loop term) should vanish as the condensate vanishes. What is
particularly significant for us is that the fixed point (\ref{fp})
is reached when $\la\bar{q}q\ra$ is dialled to zero, independently
of how the dialing is done. In the case we are concerned with here
the dialing is done by temperature (or density), so the VM fixed
point can be identified with the chiral restoration point $T=T_c$
(or $n_c$). Very near the fixed point, the constants behave simply
as
 \be
g \propto \la\bar{q}q\ra&\rightarrow& 0\, ,\nonumber\\
a-1\propto (\la\bar{q}q\ra)^2&\rightarrow& 0\,.\label{scaling1}
 \ee
This predicts the main properties for the problem at hand:
 \be
m_v \propto \la\bar{q}q\ra&\rightarrow& 0\, ,\nonumber\\
\Gamma \propto (\la\bar{q}q\ra)^2&\rightarrow& 0.\label{scaling2}
 \ee
The simple scalings (\ref{scaling1}) and (\ref{scaling2}) hold only
very near the VM fixed point. Away from the ``flash point" defined
below, the physical properties such as pole mass etc. depend on the
condensate in a much more complicated way and cannot be simply used
as a signal for chiral symmetry.
 \subsection{Violation of vector dominance (VD)}
In relativistic heavy ion collisions, there are a huge number of
sources for dileptons which need to be judiciously taken into
account for confronting experiments. For our discussion, we can
simply zero in on the $\rho$ channel which is the dominant source of
dileptons. We assume that we know how to sort out other sources, and
focus on the dipions $\pi^+\pi^-$ and the $\rho^0$ meson as the
principal dilelepton sources. Now in zero temperature and
matter-free space, vector dominance (VD) works very well, so the
dileptons are produced via
  \be
 \pi^+ + \pi^-\rightarrow \rho^0\rightarrow \gamma^*\rightarrow l^+ +l^-
  \ee
where $l=\mu$ or $e$. There is no direct $\gamma\pi^+\pi^-$  coupling
--  which is what VD means. In HLS$_1$ theory,  the photon coupling
to $\rho$ and $\pi$ is given by
 \be \delta {\cal L}= - 2eag F_\pi^2 A_{em}^\mu \Tr[\rho_\mu Q]
+2ie(1-\frac{a}{2}) A_{em}^\mu \Tr[V_\mu Q]\label{photon-coupling}
 \ee
where $A_{em}^\mu$ is the photon field, $Q$ the quark charge matrix
and $V_\mu$ the pionic vector current. The  second term describing
the photon-pion coupling vanishes when $a=2$ which is the value $a$
takes in free space, consistent with low-energy theorems. In
hot/dense matter, there are thermal/dense loop corrections, but it
is clear from (\ref{photon-coupling}) that VD must break down when
temperature/density drives $a$ to 1 as the second term starts
contributing. The first term, when coupled to $\pi^+\pi^-$, gives
rise to a vector-mediated contribution reduced from the VD value by
a factor $\sim 2$. Note that the photon-$\rho$ coupling vanishes as
$g\rightarrow 0$, which can be interpreted as the $\rho$ wave
function vanishing at the origin when the $\rho$ mass
vanishes~\cite{BHHRS}.

When one probes vector-meson properties with the dilepton as a
snap-shot, one is looking at the first term of
(\ref{photon-coupling}). While the second term as a source for
dileptons would be absent if VD held, with the violation, this is no
longer the case. The direct photon-pion coupling allows  to be produced those dileptons
that carry no direct information on the vector meson properties we
are interested in.

Medium-dependent corrections to the photon-$\rho$ coupling
$g_{\gamma\rho}=ag F_\pi^2$ can be calculated readily at one-loop
order.  The leading medium correction is the pion-loop correction to
$F_\pi^2$ and this has been computed in temperature although not in
density\footnote{Introducing density in HLS$_1$ is not yet worked
out since it would involve introducing baryons as solitons.}. The
temperature-dependent one-loop correction gives, roughly, another
factor $\sim \sqrt{2}$ to the reduction factor~\cite{HS:VD}.
Density-dependent corrections will increase this factor further.
Thus in the vicinity of the VM fixed point, there is a reduction
factor in the photon-$\rho$ coupling of $\gsim 2\sqrt{2}$ with
respect to vector dominance with $a=2$.
 \subsection{``Hadronic freedom"}
To cover the range of temperature and density involved in the
evolution of the hot/dense matter produced in heavy-ion collisions,
a useful concept is the ``flash point" at which the $\rho$ meson
recovers $\sim 90\%$ of its free-space mass, its full strong
coupling constant and $a\sim 2$.  The flash point is more or less
known in temperature, say, $T_{flash} \sim 120$ MeV but not in
density. A rough guess is that $n_{flash} \sim 2-3 n_0$. Although
density effects are not well known in HLS$_1$, there is a strong
indication that whenever nucleonic matter is present, $a$ approaches
near 1. This is observed in nucleon EM form factors. In HLS$_1$, the
baryon will appear as a skyrmion, giving a contribution to the form
factors via $V_\mu$ in (\ref{photon-coupling}) if $a\neq 2$.  Indeed, phenomenology
requires that $a$ be very near 1~\cite{HRYY-VD}, so VD is strongly violated already at one nucleon level. We expect this to
be even more so in many-nucleon systems, i.e., nuclei and nuclear
matter. This suggests that in baryonic matter between $T_c$ and
$T_{flash}$, we may safely set $a\sim 1$. This could well be an
oversimplification, so needs to be confirmed. Furthermore in this
interval, the gauge coupling is assumed to be very weak so that we can
ignore interactions involving the $\rho$ meson. This region in which
hadronic interaction is ignorable is called ``hadronic freedom"
region.

The notion of the hadronic freedom anchored on vector manifestation
of chiral symmetry has been applied to various heavy-ion processes.
It is this notion that has been invoked to argue that  all past dilepton
experiments in heavy ion collisions have failed to see the signal for BR
scaling~\cite{BHHRS}.
\section{Return of Vector Dominance}
\subsection{The problem}\label{problem}
The problem that hQCD with HLS$_\infty$ may be exploited to resolve,
assuming that it is a better approximation to QCD than HLS$_1$, can
now be precisely stated. What played a crucial role in
HLS$_1$ was that as the vector manifestation fixed point was
approached, the $\rho$ mass vanished and the parameter $a$ went to 1
while the physical pion decay constant went to zero. There both the
vanishing of the lowest-lying vector-meson mass {\em and} the {\em
violation} of VD figured importantly. The question is: Is this a
correct feature of QCD?  We would like to know what hQCD can say
about this. Let me first describe what we know about a hQCD model
that reproduces {\em correctly} certain features of chiral symmetry
of QCD predicted in the large $N_c$ and 't Hooft limit.
\subsection{HLS$_\infty$}
The notion that HLS$_1$ is an emergent gauge symmetry can be
extended to an infinite tower of flavor local fields, arriving at a
5D Yang-Mills theory called a ``dimensionally deconstructed
QCD"~\cite{Son-St}. A closely related 5D YM theory arises top-down
from string theory. A version that implements the spontaneously
broken chiral symmetry and confinement, particularly pertinent to
the problem at hand, is that constructed by Sakai and Sugimoto (SS)
exploiting D4-D8/D$\bar{8}$ branes.~\cite{SS} If I were to introduce
adequately what goes into this model -- which requires a battery of string
theory terminologies, I would have no space for what I want to discuss. I
will therefore have to skip that task entirely, referring to the
publications~\cite{SS,HRYY} for details. There are in the literature
quite a few articles that review  the SS model in detail, but
too numerous to even cite properly.

I will simply start with the 5D action derived by SS~\cite{SS} in
the form suitable for my purpose~\cite{HRYY},
  \be
S=S_{YM}+S_{CS}\label{S}
 \ee
where
 \be
S_{YM}=-\;\int dx^4 dw \;\frac{1}{4e^2(w)} \;{\Tr}
F_{MN}F^{MN}+\cdots \label{YMterm}
 \ee
where $F$ is the 5D field tensors of nonabelian flavor gauge fields,
$S_{CS}$ is the Chern-Simons action that encodes anomalies which we
won't need explicitly here and $(M,N)=0,1,2,3,4$, $w$ is the fifth
coordinate in a conformally flat coordinate and $e(w)$ is the
position-dependent ``electric coupling" of the form
\begin{equation}
\frac{1}{e^2(w)} \propto \lambda N_c M_{KK}U(w)\, .\label{ecoupling}
\end{equation}
Here $U(w)$ is the energy scale extended along the 5th ($w$)
coordinate, $\lambda$ is the 't Hooft constant $\lambda=g_{YM}^2
N_c$ and $M_{KK}$ is the Klein-Kaluza mass that sets the only scale
in the given approximation of the theory. The ellipses in
(\ref{YMterm}) stand for higher derivative terms in the expansion of
the DBI action for D8 branes in the D4 background, which are
ignored.

The important point for what follows is that the gravity action
(\ref{S}) with (\ref{YMterm}) and (\ref{ecoupling}) is dual to the
gauge theory valid in the large $\lambda$ and $N_c$ limit. With the
``probe approximation," i.e., $N_f \ll N_c$, that ignores the back
reaction of the flavor on the gluon background, the known classical
supergravity solution enters and allows one to obtain the simple
form (\ref{YMterm}). This theory may not have a correct UV completion
to QCD. But for the problem at hand, {\em what seems to crucially
matter is the generic structure of the 5D YM action that is also
arrived at bottom-up.}

In the given approximations, there are only three parameters in the
chiral limit, i.e., $\lambda$, $N_c$ and $M_{KK}$. For $N_c=3$
required by nature, both $\lambda\approx 10$ and $M_{KK}\approx
0.94$ GeV are fixed by phenomenology in the meson sector~\cite{SS}.
So there are no free parameters for the baryon sector that we are
interested in.

In order to analyze what (\ref{S}) describes in 4D, the standard
procedure is to factorize the 5th coordinate by the Kaluza-Klein
decomposition. The fifth dimension then plays the role of energy
spread in the sense of RG flow. Choosing a suitable gauge, say,
$A_5=0$, one finds that the resulting action is given by one in
which Goldstone pion fields (in the chiral limit) are coupled to an
infinite tower of nonabelian gauge fields with {\em hidden gauge
invariance}, i.e., HLS$_\infty$.
\subsection{Vector dominance in HLS$_\infty$}
This HLS$_\infty$ is found to describe surprisingly -- or rather unreasonably -- well a variety
of meson processes involving vector mesons and pions. What is most
significant -- and perhaps generic for hQCD --  is that there are no direct electroweak (EW) couplings
to the pions, rendering {\em all EW form factors of the pion totally
vector-dominated}. This can be seen immediately in the $A_5\sim \pi$
gauge. One finds, in particular, the pion charge form factor --
which is of isovector -- takes the form
 \be
F_1^\pi (Q^2)=\sum_{k=1}^\infty
\frac{g_{v^{(k)}}g_{v^{(k)}\pi\pi}}{Q^2+m_{v^{(k)}}^2}
\:\label{newvd}
 \ee
with $Q^2=-q^2$. Here the quantities  $g_{v^{(k)}\pi\pi}$ are the
trilinear couplings between pions and the vector mesons denoted
$v^{(k)}$ and $g_{v^{(k)}}$ are the photon-vector meson coupling.
Equation (\ref{newvd}) shows that in the presence of an infinite
tower, the ``old" vector dominance by the lowest $v^{(1)}=\rho$ is
replaced by a ``new" one in which the infinite tower enters.

There is no surprise in this ``new" vector dominance given that there is no direct photon-pion coupling. But what about
the photon coupling to the nucleon which we know, is not
vector-dominated when only the lowest vector mesons are present?

Here the situation looks very different at first sight because baryons
must arise as solitons. In this 5D YM theory (\ref{YMterm}), a
baryon must emerge as an instanton~\cite{HRYY,SS-baryons}. It has
been found~\cite{HRYY} that the nucleon properties that are reliably
calculable in the quenched approximation in lattice QCD simulations
can be reproduced well in this soliton model\footnote{E.g., the calculated
[experimental] values are: $g_A\approx 1.30$ [exp: 1.27] and
$(\mu_{an}^p +\mu_{an}^n)/\mu_N\approx 0$ [exp: -0.1]. The
difference from quenched lattice is expected to arise at ${\cal
O}(N_c^{-2})$ as explained in \cite{HRYY}.}.

The instanton size is found to go like $R\sim
1/M_{KK}\sqrt{\lambda}$ modulo a constant factor of order 1, so it
is pointlike in the large $\lambda$ limit. Now the minimal coupling
of EM field is holographically related to the minimal coupling of
the 5D flavor gauge field, so one expects in the action a term of
the form $\int dx^4\int dw
[-i{\B}\gamma^M(\partial_M-iA_M^{U(N)})\B$] where ${\B}$ is the 5D
baryon interpolating field for the instanton. So one would at first
sight think that there will be a coupling of the photon to the small
size instanton that is not vector-dominated in the usual sense. Such
a picture naturally arises, agreeing well with experimental data to
a large momentum transfer, in the usual Skyrme model implemented
with fluctuating vector mesons~\cite{holzwarth}, namely, the nucleon
form factor described by the photon coupling to the skyrmion and the
vector mesons, roughly of the same size as (\ref{photon-coupling})
with $a\approx 1$. This skyrmion structure had been interpreted in
terms of chiral bag model for the nucleon representing an
``intrinsic core" in which the quark degrees of freedom of QCD
reside~\cite{BRW}. The infinite tower changes this structure most
drastically. By a suitable field redefinition of the flavor gauge
fields and exploiting the RG flow in the energy spread, the direct
coupling photon-instanton coupling can be absorbed into the tower,
rendering the nucleon form factors {\em entirely} given by the
vector-dominated forms. For instance the isovector form factor is of
the form identical to that of the pion (\ref{newvd}) with only the
pion replaced by the nucleon,
 \be
F_1^N (Q^2)=\sum_{k=1}^\infty
\frac{g_{v^{(k)}}g_{v^{(k)}NN}}{Q^2+m_{v^{(k)}}^2}
\:.\label{newvd-N}
 \ee
Here the infinite tower plays an indispensable role.

There are a number of remarkable features in what we have obtained
in this model:
\begin{enumerate}
\item The infinite tower encoded in the instantonic structure
brings basic changes to the structure of elementary baryon as well
as that of dense hadronic matter. Among other things, the physical
size of the nucleon is almost entirely given by vector-meson
cloud~\cite{HRYY-VD} leaving only $\sim 0.1$ fm for the intrinsic
degree of freedom. This is in line with what is found in chiral
perturbation theory where the pion cloud plays a dominant role.
Applied to many-nucleon systems, one expects the equation of state to  be drastically
modified~\cite{MR-CND2}.
\item The vector-dominated form factors are found to work well.
Even the nucleon form factors
come out within $\sim 10\%$ accuracy for momentum transfers $Q^2\leq
1/2$ GeV$^2$.
\item
The sum rules $F_1^\pi (0)=1$ and $F_1^N (0)=1$ are both almost
completely saturated by the lowest four isovector mesons. In both
cases, the lowest $\rho$ overshoots the sum rule by $\sim 30\%$
which are mostly compensated by the next lying $\rho^\prime$.
\item
By charge conservation, one obtains a new form of universality,
$\sum_{k=1}^\infty g_{v^{(k)}\pi\pi}\simeq \sum_{k=1}^\infty
g_{v^{(k)}NN}$.
\end{enumerate}
\section{The Question for hQCD}
To address the problem posed in section \ref{problem}, we need to
understand how the masses of (at least) the low-lying four-vector
mesons and the photon coupling to them and to the baryon, change in
medium. The crucial quantity here is the quark condensate. The
efforts to introduce the quark condensate and quark masses into the
top-down hQCD -- i.e., geometrically -- are being made but the
solution remains unknown. It is known in QCD that unlike the
quantities that are well described in the large $\lambda$ and $N_c$
limit, the quark condensate is sensitive to $N_c$: For
$N_c\rightarrow \infty$, it is independent of temperature up to
$T_c$ -- which is obviously incorrect. This will be the same in
hQCD. In HLS$_1$, the violation of vector dominance and the
vanishing of the gauge coupling as given in (\ref{scaling1})
figuring crucially in hadronic freedom are closely tagged to the
quark condensate. Clearly one has to figure out how to compute
$1/N_c$ (and perhaps also $1/\lambda$) corrections.

The question to answer is: What do the HLS$_1$ properties associated with the VM mean in
terms of the infinite tower in HLS$_\infty$?

\section*{Acknowledgments}
I am grateful for a fruitful collaboration and discussions with Deog
Ki Hong, Ho-Ung Yee and Piljin Yi. I would also like to thank the
organizers of the 2008 YITP International Symposium
for the invitation to give this talk.

%

\end{document}